\documentclass[prl,aps,amssymb,twocolumn]{revtex4}
\usepackage[dvips]{epsfig,color}
\usepackage{graphicx}
\usepackage{amsmath,amsfonts,amsbsy,amssymb}
\usepackage{tabularx}

\begin{document}

\title{Split-Quaternionic Hopf Map, 
Quantum Hall Effect, and Twistor Theory}

\author{Kazuki Hasebe}
\affiliation{Department of General Education, Kagawa National College of Technology,  Takuma-cho, Mitoyo-city, Kagawa 769-1192, Japan \\
Email: hasebe@dg.kagawa-nct.ac.jp}

\begin{abstract}

Introducing a non-compact version of the Hopf map, we demonstrate remarkable close
relations between quantum Hall effect and twistor theory. We first construct
quantum Hall effect on a hyperboloid based on the noncompact 2nd Hopf map of split-quaternions.
We analyze a hyperbolic one-particle mechanics, and explore many-body
problem, where a many-body groundstate wavefunction and membrane-like excitations are derived explicitly. In the lowest Landau level, the symmetry is enhanced from $SO(3, 2)$ 
to the $SU(2, 2)$ conformal symmetry. We point out that the quantum Hall effect naturally
realizes the philosophy of twistor theory. In particular, emergence mechanism of fuzzy space-time is discussed somehow in detail.

\end{abstract}

\maketitle

 In 1967, Penrose introduced the notion of twistor \cite{TheCentralProgrammePenrose}, aiming quantization of the space-time. 
Since then, twistor has spurred the developments of  mathematical physics.  
Meanwhile, since the discovery of the quantum Hall effect (QHE) in the early 80's, 
QHE has been developed in condensed matter physics \cite{bookStone}. 
Interestingly, in the developments of the higher dimensional generalization of the QHE, their close relations have been pointed out and begun to be unveiled \cite{arXiv:hep-th/0203264,arXiv:hep-th/0210162,cond-mat/0211679,arXiv:hep-th/0307281,cond-mat/0401224}. In this paper, we further proceed to study the higher dimensional QHE and clarify common  structures between QHE and twistor theory based on a non-compact version of the Hopf map. 

In the past decade, there arose rapid developments of higher dimensional generalization of the QHE, which had been believed to be formulated only in two-dimensional spaces \cite{hep-th/0606161}. The breakthrough was brought by Zhang and Hu's four-dimensional generalization of QHE \cite{cond-mat/0110572}. Their idea was based on a mathematical concept known as the Hopf maps. The Hopf maps are mysterious topological mapping between spheres in  different dimensions, and there exist only three; 1st, 2nd and 3rd, each of which corresponds to the particular notion of the normed division algebras, $i.e.$ complex numbers, quaternions and octonions. 
As is widely known the 1st Hopf map, $S^3\overset{S^1}\longrightarrow S^2$, is the underlying mathematical structure of the Dirac monopole, and Haldane's spherical two-dimensional QHE \cite{HaldanePRL51605} owes its physical background to it.
The idea of 4D QHE is to utilize the second Hopf map, $S^7\overset{S^3}\longrightarrow S^4$.  Since the $S^3$ fibre is the group manifold of $SU(2)$, the 2nd Hopf map physically corresponds to the $SU(2)$ monopole or Yang monopole gauge field on the base-manifold $S^4$ \cite{Yang1978}. The 4D QHE represents 
  incompressible quantum liquid in such a system.  For the last 3rd Hopf map $S^{15}\overset{S^7}\longrightarrow S^8$, the corresponding monopole \cite{Grossman1984} and the 8D QHE have also been  constructed \cite{cond-mat/0306045}. Since, in the set-up of the 4D QHE, the basemanifold is $S^4$, and the $SU(2)$ monopole gauge fields are spherically symmetric, the system has the global $SO(5)$ rotational symmetry. Interestingly, the symmetry is enhanced from $SO(5)$ to $SU(4)$ in the lowest Landau level (LLL) limit, which is simply realized by taking an infinite spacing limit of Landau energy levels; $\omega=B/M$ at the ``massless limit'' ($M\rightarrow 0$).  The LLL physics of 4D QHE enjoys the $SU(4)$ symmetry, and $SU(4)$ is the Euclidean version of the $SU(2,2)$ conformal symmetry of twistor.  This ``coincidence'' implies hidden relations between the twistor theory and the QHE \cite{arXiv:hep-th/0203264,arXiv:hep-th/0210162}.   Indeed, Sparling and his collaborators analyzed 4D QHE in the formalism of the twistor theory \cite{cond-mat/0211679,cond-mat/0401224}. In particular, in Ref.\cite{cond-mat/0401224}, they suggested, if the QHE was formulated on a higher dimensional hyperboloid (ultra-hyperboloid), close structures to twistor theory would be even clearer. Independently, Karabali and Nair made use of analogies between QHE and twistor to construct the effective action for edge states \cite{arXiv:hep-th/0307281}. 

Inspired by the preceded observations, we develop a non-compact formulation of  QHE on a ultra-hyperboloid, and demonstrate remarkable close structures between twistor theory and QHE. 
For this purpose, we first explore realization of higher dimensional non-compact  Hopf maps 
 \footnote{As for the 1st Hopf map, its noncompact version is already known and the corresponding QHE on 2D-hyperboloid is constructed in Refs.\cite{arXiv:hep-th/0505095,arXiv:0809.4885}. Besides, there also exists a supersymmetric version of Hopf map \cite{LandiMarmo1981,NPB2005hasekimu} and QHE \cite{hep-th/0411137,arXiv:0809.4885}. 
}.
With ultra-hyperboloids $H^{p,q}$;  $\sum_{i=1}^p x_i^2-\sum_{j=p+1}^{p+q+1} x_{j}^2=-1$, the non-compact Hopf maps are represented as 
\begin{center}
\begin{tabular}{ccccccc}
\\ 
 & & $H^{2,1}$ &   $\overset{H^{1,0}}\longrightarrow $ & $H^{1,1}$ & & ~~~~~~~~~~(1st)\\
 &  $H^{4,3}$ & $\longrightarrow$ & $H^{2,2}$ &  & & ~~~~~~~~~~(2nd) \\
 $H^{8,7}$ & $\longrightarrow$ &   $H^{4,4}$ &&  & & ~~~~~~~~~~(3rd) \\  
\end{tabular}
\end{center}
 The construction of the non-compact version of the Hopf maps is unique;  each of them corresponds to the split-algebra, $i.e.$ split-complex numbers, split-quaternions and split-octonions \cite{prephasebe}. In this work, we utilize the non-compact 2nd Hopf map or the split-quaternionic Hopf map, $i.e.$ $H^{4,3} \rightarrow H^{2,2}$ with non-compact fibre $H^{2,1}\simeq AdS^3 \simeq SU(1,1)$. 
The total manifold $H^{4,3}$ is a hyperbola in  ``2D'' space of split-quaternions, and 
 the basemanifold $H^{2,2}$ is the split-quaternionic projective space. The $H^{2,1}$ fibre corresponds to a normalized ``1D'' split-quaternion space. To realize the 2nd non-compact Hopf map, we introduce the $(3+2)$D  $\gamma$-matrices, $\gamma^a$ $(a=1,2,3,4,5)$, which satisfy the anticommutation relations  $\{\gamma^a,\gamma^b\}=-2\eta^{ab}$ 
with $\eta^{ab}=\eta_{ab}=\text{diag}(+,+,-,-,-)$.
Their commutators yield the $SO(3,2)$ generators 
$\sigma^{ab}=-i\frac{1}{4}[\gamma^a,\gamma^b]$, 
which satisfy   
$[\sigma_{ab},\sigma_{cd}]=-i(\eta_{ac}\sigma_{bd}-\eta_{ad}\sigma_{bc}+\eta_{bd}\sigma_{ac}-\eta_{bc}\sigma_{ad}).$ 
$\gamma^a$ are explicitly given by $\gamma^i=\tau^i\otimes \sigma^2,~~\gamma^4=1\otimes \sigma^1,~~\gamma^5=-\gamma^1\gamma^2\gamma^3\gamma^4 =1\otimes \sigma^3$ ($\tau^i$ are $SU(1,1)$ generators $\tau^i=(i\sigma^1,i\sigma^2,\sigma^3)$), 
and they are skew hermitian, $(\gamma^a)^{\dagger}=-\gamma_a.$ 
The $SO(3,2)$ matrices are also represented as $\sigma_{\mu\nu}=
\begin{pmatrix}
\sigma_{\mu\nu}^{(+)} & 0 \\
0 & \sigma_{\mu\nu}^{(-)}
\end{pmatrix},$ 
where $\sigma^{(\pm)}_{\mu\nu}=\frac{1}{2}\eta_{\mu\nu i}^{(\pm)}\tau^i,$  ($\mu,\nu=1,2,3,4$) with  't Hooft ``split''-tensor 
 $\eta^{(\pm)}_{\mu\nu i}=\epsilon_{\mu\nu i }\mp \eta_{\mu i}\eta_{\nu 4}\pm \eta_{\nu i}\eta_{\mu 4},$ and  $\sigma_{\mu 5}=
\frac{1}{2}
\begin{pmatrix}
0 & \tau_{\mu} \\
 \tilde{\tau}_{\mu} & 0
\end{pmatrix}$, where $\tau_{\mu}=(\tau_i,-i)$ and 
$\tilde{\tau}_{\mu}=(\tau_i,i)$. Defining $q^i=-i\tau^i$, they satisfy the algebra of split-quaternions: $(q^1)^2=(q^2)^2=-(q^3)^2=q^1q^2q^3=1$.  
Since we are dealing with finite dimensional representation of a non-compact group $SO(3,2)$, the generators are represented by  non-hermitian matrices, 
$(\sigma^{ab})^{\dagger}=\sigma_{ab}.$ The charge conjugation matrix is constructed as $r=-\gamma^2\gamma^3=\gamma^1\gamma^4\gamma^5=
\begin{pmatrix}
\sigma^1 & 0 \\
0 & \sigma^1
\end{pmatrix}$, which has the properties; 
$r^{\dagger}=r^t=r^{-1}=r,$
$r\gamma^a r={\gamma^a}^{*},$
and $r\sigma^{ab}r=-{\sigma^{ab}}^*.$ 
The diagonalized form of $r$ is  
\begin{equation}
k=-i\gamma^1\gamma^2=i\gamma^3\gamma^4\gamma^5=
\begin{pmatrix}
\sigma^3 & 0 \\
0 & \sigma^3
\end{pmatrix},
\end{equation}
and it has the properties;  $k^{\dagger}=k^t=k^{-1}=k,$ 
$k\gamma^a k={\gamma^a}^{\dagger},$  $k\sigma^{ab}k={\sigma^{ab}}^{\dagger}.$
The hermitian matrices $k^a$ can be defined as $k^a=k\gamma^a$. Utilizing  $k^a$, 
the 2nd non-compact Hopf map is realized as 
\begin{equation}
\psi\rightarrow x^a=\psi^{\dagger}k^a\psi,
\label{2ndnoncompactHopf}
\end{equation}
where $\psi$, which we call the non-compact 2nd Hopf spinor, is a  $SO(3,2)$ Dirac spinor subject to a normalization condition; $\psi^{\dagger} k\psi= 1,$ 
and then, regarded as coordinates on $H^{4,3}$. 
Since $k^a$ are hermitian matrices,  $x^a$ given by (\ref{2ndnoncompactHopf}) are real, and satisfy the condition,  
$\eta_{ab}x^a x^b =-(\psi^{\dagger}k\psi)^2=-1$, which defines  $H^{2,2}$. 
Inverting the 2nd non-compact Hopf map, the non-compact 2nd Hopf spinor is represented as 
\begin{equation}
\psi=\frac{1}{\sqrt{2(1+x^5)}}
\begin{pmatrix}
(1+x^5)\phi\\
(x^4-ix^i\tau_i)\phi
\end{pmatrix},
\label{explicit2ndspinor}
\end{equation}
where  $\phi$ denotes an arbitrary complex two-component spinor subject to a normalization condition 
$\phi^{\dagger}\sigma^3\phi=1$, representing the $AdS^3$-fibre.
The associated canonical connection is induced as $A=-i\psi^{\dagger} k d\psi=dx^a \phi^{\dagger}\sigma^3 A_a\phi$, where $A_a$ is given by  
$A_{\mu}=-\sigma_{\mu\nu}^{(+)}\frac{x^{\nu}}{1+x^5} =-\frac{1}{2}\eta_{\mu\nu i }^{(+)} \frac{x^{\nu}}{1+x^5}\tau^i,~
A_5=0$ (hereafter, we omit $(+)$ on $\sigma_{\mu\nu}$ and $\eta_{\mu\nu i}$), 
which are naturally regarded as $SU(1,1)$ non-abelian monopole gauge field. The corresponding $SU(1,1)$ field strength, 
$F_{ab}=\partial_a A_b-\partial_b A_a+i[A_a,A_b]$, is evaluated as   
$F_{\mu\nu}= x_{\mu}A_{\nu}-x_{\nu}A_{\mu}+\sigma_{\mu\nu}$, $F_{\mu 5} = ( 1+x^5)A_{\mu}.$ 
Thus, the non-compact 2nd Hopf map physically corresponds to a set-up of 4D hyperboloid $H^{2,2}$ in  $SU(1,1)$ monopole background. 

We first analyze Landau problem in such a system. (Similar but another hyperbolic Landau problem has been discussed in Ref.\cite{arXiv:hep-th/0602231}). The $SO(3,2)$ covariant angular momentum is defined as $\Lambda_{ab}= -ix_a D_b+ix_b D_a$ where $D_a=\partial_a+iA_a.$
The covariant angular momentum satisfies the relation,  
$[\Lambda_{ab},\Lambda_{cd}]=i(\eta_{ac}\Lambda_{bd}-\eta_{ad}\Lambda_{bc}+\eta_{bd}\Lambda_{ac}-\eta_{bc}\Lambda_{ad})
-i(x_a x_c F_{bd}-x_ax_dF_{bc}+x_bx_dF_{ac}-x_bx_cF_{ad}).$
The total angular momentum is constructed as $L_{ab}=\Lambda_{ab}-F_{ab}$
and generates the $SO(3,2)$ transformation;  
$[L_{ab},T_{cd}]=i(\eta_{ac}T_{bd}-\eta_{ad}T_{bc}+\eta_{bd}T_{ac}-\eta_{bc}T_{ad})$, with  $T_{ab}=L_{ab},\Lambda_{ab}$ and $F_{ab}$. Especially, when $T_{ab}=L_{ab}$, the algebra represents the closed $SO(3,2)$ algebra of  $L_{ab}$.  
The one-particle Landau Hamiltonian is given by  
$H=-\frac{1}{2M}\eta_{ab}D^a D^b$ 
where   
$\eta_{ab}D^a D^b=-\frac{\partial^2}{\partial R^2}-(d-1)\frac{1}{R}\frac{\partial}{\partial R}+\frac{1}{R^2}\sum_{a<b}\Lambda_{ab}^2$
 ($d=5$).  
Here,  $R$ denotes the radial coordinate given by  $\eta_{ab}x^a x^b=-R^2$. 
On the surface of $H^{2,2}$, the Landau Hamiltonian is reduced to 
\begin{equation}
H=-\frac{1}{2MR^2}\sum_{a<b} \Lambda_{ab}^2.
\label{oneparticlehamiltonian}
\end{equation}
The covariant angular momentum is orthogonal to the field strength  $\Lambda_{ab}F^{ab}=F_{ab}\Lambda^{ab}=0$, and the Hamiltonian is rewritten as 
$H=-\frac{1}{2MR^2}\sum_{a<b}(L_{ab}^2-F_{ab}^2)$. 
The eigenvalue of the $SO(3,2)$ Casimir operator is  
$C=\sum_{a<b}L_{ab}^2=\mathcal{E}(\mathcal{E}-3)+s(s+1)$ 
with $\mathcal{E}=-s-n$ ($n=0,1,2,\cdots$ and $s=0,-\frac{1}{2},-1,-\frac{3}{2},\cdots$) \cite{LoeweLecNote,EvansJMP1967}.
Meanwhile, 
$\sum_{a<b}F_{ab}^2=2s(s+1)$
where 
$s=-\frac{I}{2}$  
with $SU(1,1)$ monopole charge $\frac{I}{2}=0,\frac{1}{2},1,\cdots$. 
Then, for discrete series of the $SO(3,2)$ group, the energy eigenvalue of (\ref{oneparticlehamiltonian}) reads as  
$E_n=\frac{1}{2MR^2}(I(n+1)-n(n+3)),$
where $n$ represents Landau level index. 
The discrete spectrum takes a form of an upper convex, and is not unbounded below. However, the LLL ($n=0$) with energy $E_{LLL}=\frac{I}{2MR^2}$ is not completely unstable but meta-stable, since there exists a ``potential barrier'' between the LLL and the negative energy levels.  In the thermodynamic limit: $R,I\rightarrow \infty$ with magnetic length $\ell_B=R\sqrt{\frac{2}{I}}$ fixed, the potential barrier becomes larger and the LLL becomes stabler. There also exists continuous spectrum, but it does not contribute to Landau levels in the thermodynamic limit, since it specifies energy spectrum higher than the discrete energy levels and behaves as $\sim \frac{1}{2MR^2}((\frac{I}{2})^2+\nu^2)$ ($\nu$ is the continuous parameter) in the limit.   
Indeed,  the planar Landau level $\frac{I}{2MR^2}(n+1)$ can be fully reproduced only by the discrete spectrum in the limit. The above behaviors of the $SO(3,2)$ Landau problem are quite analogous to those of the $SU(1,1)$ Landau problem \cite{Comtet1987} because of the similar group structures between $SU(1,1)$ and $SO(3,2)$, $i.e.$ $Sp(2,R)\simeq SU(1,1)$ and $Sp(4,R)\simeq SO(3,2)$. 

Next, we discuss many-body problem on $H^{2,2}$.  
In the original spherical 2D QHE, the Laughlin-Haldane groundstate wavefunction is constructed by a  $SU(2)$ singlet combination of the 1st Hopf spinors \cite{HaldanePRL51605}. Thus, the Laughlin-Haldane wavefunction respects the isometry of the basemanifold, namely, $SO(3)$ symmetry of $S^2$. Physically, the symmetry expresses uniform distribution of the ground state quantum liquid  on the surface of $S^2$. In the present, the basemanifold is $H^{2,2}$ whose isometry is $SO(3,2)$, so it might be reasonable to adopt a $SO(3,2)$ singlet wavefunction made by the 2nd  non-compact Hopf spinors as  the groundstate wavefunction.
The charge conjugation of $SO(3,2)$ spinor $\psi$ is  constructed as 
$\psi^c=r\psi^*$, and, without introducing complex variables, $SO(3,2)$ singlet wavefunction can be constructed as 
\begin{equation}
\Psi=\prod_{i<j} (\psi^t_i  rk \psi_j)^m,
\label{SO(32)LaughlinHaldane}
\end{equation}
which we adopt as the higher dimensional analogue of the Laughlin-Haldane wavefunction. 
The wavefunctions for topological excitation can also be derived by following the procedure given by Haldane \cite{HaldanePRL51605}.
The topological excitations are induced by flux penetrations, and 
their annihilation and creation operators are, respectively, given by 
\begin{equation}
A(\chi)=\prod_i^N \chi^{\dagger}r \frac{\partial}{\partial\psi_i},~~~~A^{\dagger}(\chi)=\prod_i^N \psi^t_i  rk \chi,
\label{membraneexcitation}
\end{equation}
where $\chi$ denotes a flux penetration point on $H^{2,2}$ by the relation $\chi^{\dagger} k^a\chi=\Omega^a(\chi)$. 
Indeed, the operators (\ref{membraneexcitation}) satisfy the creation and annihilation relations, $[A(\chi),A^{\dagger}(\chi)]=1,~[A(\chi),A(\chi')]=0$, and 
$[A^{\dagger}(\chi),A^{\dagger}(\chi')]=0.$ With fuzzy hyperboloid coordinates $X_a=-\psi^t \gamma_a^t\frac{\partial}{\partial\psi}$ (its derivation will be discussed later), the creation operator satisfies  
$[\Omega_a(\chi)X^a, A^{\dagger}(\chi)]=N A^{\dagger}(\chi). $
This implies that  $N$-particles on $H^{2,2}$ are pushed ``outwards'' from the point of flux penetration, and a charge deficit is generated at the  point.   It is noted that $\chi$
 carries ``extra degrees'' of $AdS^3$-fibre except for the degrees denoting the point on $H^{2,2}$,  and, up to $U(1)$ phase, such extra degrees account for membrane of the form $H^{2,0} \simeq AdS^3/U(1)$. 
Thus, though the topological excitations are  point-like on $H^{2,2}$, they carry  membrane-like internal structures. 

To clarify analogies between QHE and twistor theory, we exploit the Lagrange formalism. 
Lagrangian of one-particle mechanics is given by 
\begin{equation}
L=\frac{M}{2}\eta_{ab}\dot{x}^a\dot{x}^b+ \dot{x}^a A_a,
\end{equation}
where $A=dx^a A_a= -i\psi^{\dagger}k d\psi$. Since the particle is confined on a surface of  $H^{2,2}$, a constraint should be imposed on $x^a$; $\eta_{ab}x^ax^b=1$. (For simplicity, we take $R=1$ hereafter.)
Apparently, the Lagrangian and the constraint respect  the $SO(3,2)$ symmetry. Meanwhile in the LLL limit $M\rightarrow 0$, the kinetic term drops, and the gauge interaction term only survives to yield,  
$L_{LLL}= \dot{x}^a A_a= -iI\psi^{\dagger}k \frac{d\psi}{dt},$
with the constraint $\psi^{\dagger}k\psi=1$.
For later convenience, we scale the Hopf spinor as $\psi \rightarrow \frac{1}{\sqrt{I}}\psi$, and  the LLL Lagrangian is written as 
\begin{equation}
L_{LLL} =-i\psi^{\dagger}k \frac{d\psi}{dt},
\label{LLLaction}
\end{equation}
and the constraint as 
\begin{equation}
\psi^{\dagger}k\psi=I.
\label{LLLconstraint}
\end{equation}
One may notice that both the LLL Lagrangian (\ref{LLLaction}) and the constraint (\ref{LLLconstraint}) respect the $SU(2,2)$ conformal symmetry.
Here, we invoke the twistor description of a massless particle based on Ref.\cite{Shirafujiptp7018}. 
The momentum of free massless particle satisfies the relation $\xi_{\mu\nu}p^{\mu}p^{\nu}=0$ ($\xi_{\mu\nu}$ is the Lorentzian metric: $\xi_{\mu\nu}=diag(+,+,+,-)$),  and can be expressed as $p^{\mu}=\pi^{\dagger}\sigma^{\mu}\pi$ with arbitrary two-component $SL(2,C)$ spinor $\pi_{\alpha}$. Twistors are a $SU(2,2)$ four-component representation $Z^a=(Z^1,Z^2,Z^3,Z^4)$, where the lower two-components $Z^3$ and $Z^4$ are given by $(Z^3,Z^4)=(\pi_1,\pi_2)$ and the upper components $Z^1$ and $Z^2$ are introduced as 
\begin{equation}
\begin{pmatrix}
Z^1\\
Z^2
\end{pmatrix}
=ix^{\mu}_M\sigma_{\mu}
\begin{pmatrix}
Z^3\\
Z^4
\end{pmatrix}.
\label{originaltwistorincidence}
\end{equation}
(The repeated indices $\mu$ here are contracted by Lorentzian metric). Eq.(\ref{originaltwistorincidence}) plays a central role in twistor theory, and is known as the incidence relation that represents relations between original Minkowski space-time and twistor space. Meanwhile, eliminating the $AdS^3$ gauge freedom $\phi$ in (\ref{explicit2ndspinor}), one may derive the following relation between the upper and lower two-components of the Hopf spinor:   
\begin{equation}
\begin{pmatrix}
\psi^1\\
\psi^2
\end{pmatrix}
= i{x}^{\mu}_L\tau_{\mu}
\begin{pmatrix}
\psi^3\\
\psi^4
\end{pmatrix}, 
\label{incidenceHopfspinor}
\end{equation}
where ${x}^{\mu}_L$ denotes the stereographic coordinates on the four-dimensional Lobachevsky  plane ${x}^{\mu}_L\equiv\frac{1}{1-x^5}x^{\mu}$. 
Eq.(\ref{incidenceHopfspinor}) expresses relations between coordinates in the hyperbolic manifolds, $H^{2,2}$ and $H^{4,3}$, and Eq.(\ref{incidenceHopfspinor}) may be regarded as the incidence relation in the version of the non-compact QHE. Analogies between the two incidence relations (\ref{originaltwistorincidence}) and (\ref{incidenceHopfspinor}) are apparent, and their correspondence reads as     
\begin{equation}
({x}^{1}_M,{x}^2_M,{x}^3_M,{x}^0_M)\leftrightarrow ({x}^1_L,{x}^2_L,i{x}^3_L,{x}^4_L). 
\end{equation}
The imaginary factor in front of ${x}^3_L$ stems from the signature difference of their metrics; $(+,+,+,-)$ and  $(+,+,-,-)$. 
With use of twistors, 
the massless particle Lagrangian is simply written as
$L=-iZ_a^*\frac{d}{d\tau}Z^a$ \cite{Shirafujiptp7018}, 
where $Z_a$ is the dual twistor $Z_a=(\pi_{\alpha},\omega^{\beta})$ and $\tau$ the invariant time. 
 With ``diagonalized'' twistors  $(\mathbb{Z}^1,\mathbb{Z}^2,\mathbb{Z}^3,\mathbb{Z}^4)=\frac{1}{\sqrt{2}}(Z^1+Z^3,Z^1-Z^3,Z^2+Z^4,Z^2-Z^4)$, the twistor Lagrangian becomes 
\begin{equation}
L=-i\mathbb{Z}^{\dagger} k\frac{d}{d\tau}\mathbb{Z}.
\label{twistoraction}
\end{equation}
The norm of $\mathbb{Z}$ corresponds to the helicity of massless particle; 
\begin{equation}
\mathbb{Z}^{\dagger}k \mathbb{Z}=2\lambda.
\label{normtwistor}
\end{equation}
(After quantization, the helicity $\lambda$ takes an integer or half integer.) 
In a massless limit of free particle, the system enjoys the $SU(2,2)$  conformal symmetry rather than the Poincare symmetry, and such $SU(2,2)$ symmetry is manifest in both (\ref{twistoraction}) 
and (\ref{normtwistor}). Now, one may observe striking analogies between the non-compact QHE and twistor; as for their actions (\ref{LLLaction}) and (\ref{twistoraction}), as well as their constraints (\ref{LLLconstraint}) and (\ref{normtwistor}) [See Table \ref{tableanalogies} also.]  
\begin{table}
\renewcommand{\arraystretch}{1}
\hspace{5mm}
\begin{tabular}{|c|c|c|}
\hline    &  QHE & ~~~~~~ Twistor ~~~~ 
\\
\hline\hline Fundamental quantity  &  Hopf spinor
& Twistor \\ 
\hline Quantized value  & Monopole charge & Helicity \\
\hline
Base manifold & Hyperboloid $H^{2,2}$  &   Minkowski space
\\
 \hline
Original symmetry &  $SO(3,2)$ &  Poincare 
\\ 
\hline
Special limit & LLL ($M\rightarrow 0$) & zero mass ($M\rightarrow 0$)  \\
\hline
 Enhanced symmetry & $SU(2,2)$ & $SU(2,2)$ \\
\hline Emergent Manifold  & $CP^{3}$
& $CP^{3}$ \\
\hline Fuzzy manifold  & Fuzzy hyperboloid
& Fuzzy twistor space\\
\hline
\end{tabular}
\caption{Analogies between the non-compact QHE and twistor: The original set-ups are different; the basemanifold of the QHE is $H^{2,2}$ whose isometry
 is $SO(3,2)$, while that of twistor is Minkowski space whose isometry is
  Poincare. However, once ``massless limit'' is taken, both systems 
 enjoy the enlarged $SU(2,2)$ conformal symmetry and everything goes parallel. 
[See also Ref.\cite{arXiv:hep-th/0004032}, in which twistor formalism was applied to describe a charged particle in monopole background.]  } 
 \label{tableanalogies}
\end{table} 

 We proceed to the quantization of the LLL, and see how the
 QHE naturally realizes the original philosophy of the twistor theory. 
From (\ref{LLLaction}), the canonical conjugate variable of $\psi$ is derived as $\pi=-i\psi^{\dagger}k$; $\pi$ is not the time derivative of $\psi$, but its complex conjugation.
As is well known, this brings the emergence of 
non-commutative geometry.
The quantization condition is imposed between $\psi$ and $\psi^*$,
so $\psi^*$ is regarded as 
\begin{equation}
\psi^*=-k\frac{\partial}{\partial\psi}.
\label{derivativeandcomplex} 
\end{equation}
In quantum mechanics, the constraint (\ref{LLLconstraint}) is transformed to a condition on LLL states, and the LLL states are constructed by the basis 
$\psi_{m_1,m_2,m_3,m_4}=\psi^{m_1}\psi^{m_2}\psi^{m_3}\psi^{m_4}$, which is a symmetric tensor product of the non-compact Hopf spinors,  
with   
$m_1+m_2+m_3+m_4=I.$ Substituting (\ref{derivativeandcomplex}) to the expression of $x^a$ (\ref{2ndnoncompactHopf}), we now see that $x_a$ is regarded as the operator 
\begin{equation}
X_a=-\psi^t\gamma_a^t\frac{\partial}{\partial\psi}.
\end{equation}
Meanwhile, in LLL, the $SO(3,2)$ generator $L_{ab}$ is effectively represented as 
\begin{equation}
X_{ab}=-\psi^t\sigma_{ab}^t\frac{\partial}{\partial \psi},
\label{identifLLL}
\end{equation}
since $L_{ab}\psi=-\sigma_{ab}\psi$.  $X^a$ and  $X^{ab}$ satisfy the following algebra;   
\begin{align}
&[X_a,X_b]=4iX_{ab},~~[X_a,X_{bc}]=i(\eta_{ab}X_c-\eta_{ac}X_b),\nonumber\\
&[X_{ab},X_{cd}]=-i(\eta_{ac}X_{bd}-\eta_{ad}X_{bc}+\eta_{bd}X_{ac}-\eta_{bc}X_{ad}).  
\label{noncommubetxx}
\end{align}
\vspace{-0.3cm}

With definition $X_{AB}$ $(A,B=1,2,\cdots, 6)$; $X_{a6}=-\frac{1}{2}X_a$ and $X_{ab}=X_{ab}$, Eq.(\ref{noncommubetxx}) represent the $SO(4,2)\simeq SU(2,2)$ algebra of $X_{AB}$. It is worthwhile to notice that $X^a$ do not form a closed algebra by themselves, but form a closed algebra if  $X_{ab}$ are introduced. The basic notion of non-commutative geometry is ``algebraic construction of geometry''.
The ``unclosed algebra'' of $X^a$ suggests that the fuzzy geometry of $H^{2,2}$ cannot be constructed solely by $X^a$, but  ``demands'' an extra space spanned by $X_{ab}$  
\footnote{Such ``enhancement'' mechanism is first reported in the context of  the fuzzy $S^4$ \cite{HoRamgoolamNPB2002,Kimura2002}. 
In Ref.\cite{HoRamgoolamNPB2002,Kimura2002}, it is claimed that the fuzzy geometry of $S^4$ is specified by the symmetric tensor product representation of the $SO(5)$ gamma matrices. 
In the present non-compact case, the fuzzy geometry of $H^{2,2}$ is similarly specified by the symmetric tensor product representation of the non-hermitian $SO(3,2)$ gamma matrices.}. 
The $SU(2,2)$ non-commutative algebra naturally defines the fuzzy manifold of  $CP^{2,1}$, which is the projective twistor space locally equivalent to $H^{2,2}\times H^{2,0}$. Thus, the corresponding fuzzy manifold of $H^{2,2}$ is $\it{not}$ a 4D  but a 6D manifold, and 
the extra $H^{2,0}$-space is the very space induced by the requirement of  the non-commutative geometry. 
Consequently, the fuzzy $H^{2,2}$ may be given by 
\begin{equation}
H^{2,2}_F\simeq SO(3,2)/U(1,1),
\end{equation}
which is topologically equivalent to $CP^{2,1}$. 
Here, we add some crucial comments.  To derive the non-commutative algebras  (\ref{noncommubetxx}), we did $\it{not}$ quantize the original space-time coordinates by themselves, but quantized the more fundamental (Hopf spinor) variables, and the fuzziness in the original space-time was induced by that of the more fundamental space.  Indeed, this realizes the original philosophy of twistor; the space-time fuzziness should come from the more fundamental (twistor) space \cite{TheCentralProgrammePenrose}!   
The non-commutative geometry is deeply related to particular physics in QHE.  
In the LLL ($M\rightarrow 0$), the covariant angular momentum drops to yield  $L_{ab}\rightarrow -F_{ab}$, and the non-commutative relation of $X_a$ will be given by $[X_a,X_b]=i\frac{1}{4}F_{ab}$. Then, the equation of motion is derived as 
\begin{equation}
I_a=\dot{X}_a=-i[X_a,V]=-\frac{1}{4}F_{ab}E^b
\end{equation}
where $E_a=-\partial_a V$, and  the Hall effect, $I^a E_a=0$, is confirmed.
 Around the north pole, non-commutative relation becomes 
\begin{equation}
[{X}_{\mu},X_{\nu}]=  i \ell_B^2 \eta_{\mu\nu i} \tau^i.
\end{equation}
This is the fundamental relation for the split-quaternionic geometry unifying the space-time fuzziness and the internal ``spin'' structure, as first pointed in the original set-up of the 4D QHE \cite{cond-mat/0110572}.  

To summarize, having exploited the non-compact version of the 2nd Hopf map, we 
clarified close mathematical and physical structures between  QHE and twistor theory. 
Moreover, based on the non-commutative geometry arguments, it was shown that the QHE naturally realizes the original philosophy of twistor theory.
We also explored Landau problem on $H^{2,2}$ and many-body physics where higher dimensional analogues of quantum liquid and topological excitations were derived explicitly. 

The non-compact QHE owes its mathematical background to the non-compact Hopf map. 
 A particular feature of such Hopf-map-based construction would be uniqueness: the space-time manifold, gauge symmetry, global symmetry are uniquely determined by the geometrical structure of the Hopf map. 
At the same time,  due to the usage of its non-compact version, there arises an exotic feature: extra-time dimensions. 
Indeed, the present basemanifold $H^{2,2}$ has two-temporal dimensions as well as two spacial dimensions. 
Extra time physics has been discussed in various contexts [See for instance, Refs.\cite{hep-th/0502065,Toppan2001,SparlingTillmancond-mat/0401015,Hul19982000}], and  
the present model might demonstrate particular properties speculated in  extra time physics. 
Especially, analogies to Bars' 2T physics \cite{hep-th/0502065} are  quite suggestive: In 2T physics, the (enhanced) global symmetry is $SU(2,2)$ and the gauge symmetry is $Sp(2,R)\simeq SU(1,1)$ which is crucial to eliminate negative norm states. Interestingly, also in the present model, the (enhanced) global symmetry is $SU(2,2)$ and the gauge symmetry is $SU(1,1)$, which is automatically  incorporated by the geometry of the non-compact 2nd Hopf map. 
This seems to suggest hidden relations between the 2T physics and the present model. 
The edge excitations are also worthwhile to be investigated. As edge excitations, the original 4D QHE exhibits  higher spin massless spectrum including photon and graviton \cite{cond-mat/0110572}. However, in flat space-time, a field theoretical description of higher spin massless particles has not successfully been constructed. Meanwhile, in $AdS$ space with negative curvature, a consistent formulation of higher spin field theory is possible \cite{hep-th/9910096}. Fortunately, the present basemanifold is hyperbolic and its edge manifold also possesses negative curvature.  Then, it is expected that the present edge model could yield a consistent higher spin theory in negative curvature space. 

In the set-up of the non-compact 4D QHE, we have encountered diverse novel mathematics and physical ideas, such as split-quaternions, non-compact Hopf map, non-commutative geometry, twistor theory, higher spin theory, and even extra-time physics. Such ``richness'' may imply profound structures behind the present construction. 

\vspace{0.2cm}
The author would like to express his deep gratitude to Prof. S.C. Zhang for bringing author's attention to the present subject.  
 The author also wish to thank 
 Dr. Y. Kimura, Prof.  J. Van der Jeugt, H. Kunitomo, R. Sasaki and K. Takasaki for valuable discussions. 
This work was supported by Sumitomo Foundation.

\vspace{-0.6cm}

\end{document}